\documentclass[journal=jpclcd,manuscript=letter]{achemso}

\usepackage{mathrsfs}
\usepackage{amsmath}
\usepackage{rotating}
\usepackage{amssymb}
\usepackage{graphicx}
\usepackage{bm}
\usepackage[colorlinks=true,linkcolor=blue]{hyperref}

\usepackage{siunitx}
\DeclareSIUnit\amagat{amg}
\DeclareSIUnit\molar{M}

\usepackage{color}

\definecolor{myviolet}{rgb}{0.5,0,0.5} 
\definecolor{myred}{rgb}{1,0,0} 
\definecolor{myorange}{rgb}{1,.5,0} 
\definecolor{mygreen}{rgb}{0,0.5,0} 
\definecolor{mybrown}{cmyk}{0,0.50,1,0.41}\definecolor{myblue}{rgb}{0,0,0.75} 
\definecolor{mymagenta}{cmyk}{0,1,0,0.12}

\expandafter\ifx\csname package@font\endcsname\relax\else
\expandafter\expandafter
\expandafter\usepackage
\expandafter\expandafter
\expandafter{\csname package@font\endcsname}
\fi

\title{Real-Time Polarimetry of Hyperpolarized \textsuperscript{13}C Nuclear Spins using an Atomic Magnetometer}

\author{Kostas Mouloudakis}
\affiliation{ICFO -- Institut de Ci\`encies Fot\`oniques, The Barcelona Institute of Science and Technology, 08860 Castelldefels (Barcelona), Spain}
\author{Sven Bodenstedt}
\affiliation{ICFO -- Institut de Ci\`encies Fot\`oniques, The Barcelona Institute of Science and Technology, 08860 Castelldefels (Barcelona), Spain}
\author{Marc Azagra}
\affiliation{IBEC -- Institute for Bioengineering of Catalonia, 08028 Barcelona, Spain}
\author{Morgan W. Mitchell}
\affiliation{ICFO -- Institut de Ci\`encies Fot\`oniques, The Barcelona Institute of Science and Technology, 08860 Castelldefels (Barcelona), Spain}
\affiliation{ICREA -- Instituci\'{o} Catalana de Recerca i Estudis Avan\c{c}ats, 08010 Barcelona, Spain}
\author{Irene Marco-Rius}
\affiliation{IBEC -- Institute for Bioengineering of Catalonia, 08028 Barcelona, Spain}
\author{Michael C. D. Tayler}
\email{michael.tayler@icfo.eu}
\affiliation{ICFO -- Institut de Ci\`encies Fot\`oniques, The Barcelona Institute of Science and Technology, 08860 Castelldefels (Barcelona), Spain}

\date{January 19\textsuperscript{th}, 2023 [revision 2]}%


\begin{document}

\begin{tocentry}
\includegraphics[width=2in]{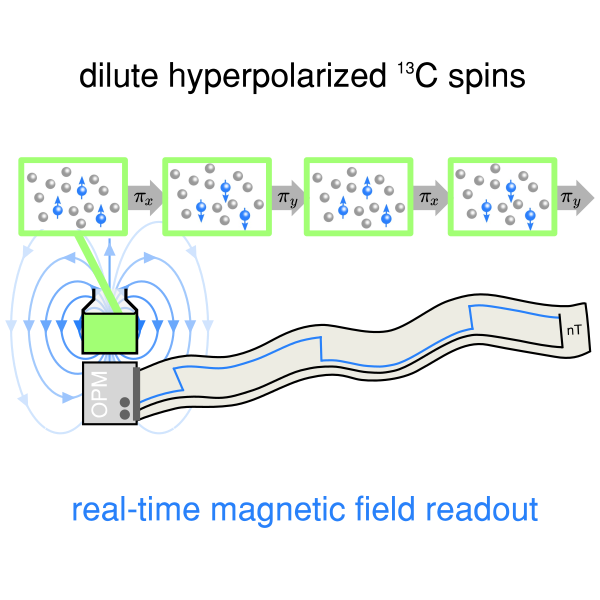}
\end{tocentry}

\begin{abstract}
\noindent We introduce a method for non-destructive quantification of nuclear spin polarization, of relevance to hyperpolarized spin tracers widely used in magnetic resonance from spectroscopy to in vivo imaging.  In a bias field of around \SI{30}{\nano\tesla} we use a high-sensitivity miniaturized \textsuperscript{87}Rb vapor magnetometer to measure the field generated by the sample, as it is driven by a windowed dynamical decoupling pulse sequence that both maximizes the nuclear spin lifetime and modulates the polarization for easy detection. We demonstrate the procedure applied to a \SI{0.08}{\molar} hyperpolarized [1--\textsuperscript{13}C]-pyruvate solution produced by dissolution dynamic nuclear polarization, measuring polarization repeatedly during natural decay at Earth's field.  Application to real-time quality monitoring of hyperpolarized substances is discussed. 
\end{abstract}

\noindent Magnetic resonance spectroscopy (NMR) and imaging (MRI) of nearly fully polarized (also known as ``hyperpolarized'') nuclear-spin ensembles is an emerging methodology\cite{Aquaro2014,Comment2014Biochem,Brindle2015,Wang2019,Jrgensen2022} for probing cellular biochemistry 
in vitro\cite{}, 
in vivo\cite{MarcoRius2017emagres,MarcoRius2020MRMPBM,vonMorze2017MRM,Can2022CommBio} and in perfused organs \cite{Merritt2007PNAS}.   
The magnetic moment of a hyperpolarized substance, which is proportional to the nuclear polarization, can be three to four orders of magnitude larger than that of thermally polarized substances\cite{ArdenkjrLarsen2003PNAS}, resulting in MRI pixels that contain chemically specific signals high above the noise floor even at low concentration. 
In pre-clinical\cite{Ros2021STAR} and clinical\cite{Gallagher2020PNAS,Woitek2021BJC} monitoring of metabolic diseases, hyperpolarized small molecules such as pyruvic acid are typically injected into a biosystem where metabolism takes place.  
Dynamic MRI of the downstream metabolites then provides biodistribution maps and kinetic rates to distinguish between healthy and diseased tissue.

In these and other applications, such as noble-gas MRI, there are several motives to quantify nuclear spin hyperpolarization.  The magnitude of polarization, $P$ (a dimensionless number ranging from 0, for unpolarized spins, to $\pm1$, when spins are completely polarized), determines the effective concentration of the substrate seen by MRI and therefore is a crucial parameter in kinetic models of spin-label transfer along a metabolic pathway, especially in cases where hyperpolarized agents are administered to the subject repeatedly\cite{Lees2020NMRBiomedicine}.  Additionally, any exogenous substance introduced in vivo is subject to strict quality control and dose requirements.  Given that many hyperpolarization technologies are at an early development stage and not always consistent, $P$ is currently a control variable; for instance, in dissolution dynamic nuclear polarization, polarization is monitored both before\cite{Elliott2021,Koptyug2022PCCP24} and after\cite{ArdenkjaerLarsen2011NMRBiomed24,Kurhanewicz2019Neoplasia} sample dissolution.
A reliable measurement of $P$ may also provide a cost-per-unit signal enhancement metric to compare different hyperpolarization methods, which may be influential in future large-scale clinical or other applicability of hyperpolarized magnetic resonance.  

A standard polarimetry method is \textit{ex post} comparison of the hyperpolarized-NMR signal amplitude against the NMR signal of the same system after polarization has decayed back to a thermal equilibrium value\cite{ArdenkjaerLarsen2011NMRBiomed24,Kurhanewicz2019Neoplasia}.  This approach can be performed on a variety of NMR spectrometer systems in background fields from mT to T, but is slow and requires several calibrations such as tuning of radiofrequency coils, pulse flip angles, and magnetic field; conventionally, although lower-field instrumentation may be less expensive, calibrations become much lengthier due to reduced-efficiency inductive pickup of the spin precession signals\cite{Korchak2012AMR44}.  In special molecules one can also determine polarization from intensity patterns in spin-coupling multiplets, namely where the hyperpolarized substrate contains a second, ancillary spin label to be used as a ``spy'' nucleus, e.g., [1,2-\textsuperscript{13}C\textsubscript{2}]-pyruvate\cite{MarcoRius2013,IMRPhDthesis,Vuichoud2015JMR260} or [2-\textsuperscript{13}C]-acetate\cite{Elliott2021}.  
A third method is through detection of the bulk magnetic field, $\mathbf{B}$, produced by the hyperpolarized spins.  For an ensemble of uncoupled spin-1/2 nuclei,
\begin{equation}
  |\mathbf{B}| = \hbar\mu_0 P C \gamma k\,,
  \label{eq:Mz}
\end{equation}
assuming that $\mathbf{B}$ is parallel to the polarization vector, where $\hbar$ is the reduced Planck constant, $\mu_0$ is the vacuum permeability, $C$ is the number density of spins, $\gamma$ is the gyromagnetic ratio and $k$ is a constant of proportionality that depends on the geometry of the sample and the position where $\mathbf{B}$ is measured. 
In the case of a uniformly polarized spherical sample of radius $R$, $k=(R/r)^3/3$ for position $r>R$ along the $\mathbf{B}$ direction.

Most bulk-protonated liquids have $|\mathbf{B}|$ on the order of \si{\pico\tesla} to \si{\nano\tesla} for $R/r \approx 1$, at thermal equilibrium in moderate fields, e.g., 1 T. 
A dilute hyperpolarized liquid ($PC \approx P_{\rm thermal}C_{\rm bulk}$) could produce fields of comparable magnitude, while a hyperpolarized neat liquid ($C \approx C_{\rm bulk}$, $P\gg P_{\rm thermal}$) will produce a considerably stronger field.
One way to measure $|\mathbf{B}|$ is through Zeeman frequency shifts of a second, thermally polarized spin species in the liquid\cite{Eichhorn2022jacs}. This, in addition to the above methods, however, relies on high-resolution NMR instrumentation (with which comes high cost and limited mobility) and the destruction of a significant part if not all of the hyperpolarized spin order.  In the following, as an alternative, we demonstrate a direct, nondestructive measurement of $|\mathbf{B}|$ using an optically pumped magnetometer (OPM)\cite{Savukov2005PRL,Tayler2017RSI,Yashchuk2004PRL93}.  To obtain $P$, the factor $k$ in \autoref{eq:Mz} is eliminated using a reference sample of known polarization ($P_{\rm ref}$) and concentration ($C_{\rm ref}$): 
\begin{equation}
  P = \left( \frac{\gamma_{\rm ref}}{\gamma} \right) \left( \frac{C_{\rm ref}}{C} \right) \frac{|\mathbf{B}|}{|\mathbf{B}_{\rm ref}|}P_{\rm ref}\,.
  \label{eq:PA}
\end{equation}
Here the reference sample is \textsuperscript{1}H spins in thermally polarized non-degassed milli-Q water at \SI{2}{\tesla} (\textsuperscript{1}H\textsubscript{2}O; $C_{\rm ref} = $ 110 \si{\molar}, $P_{\rm ref} \approx 6.8 \times 10^{-6}$).  

\begin{figure}%
\includegraphics[width=0.75\columnwidth]{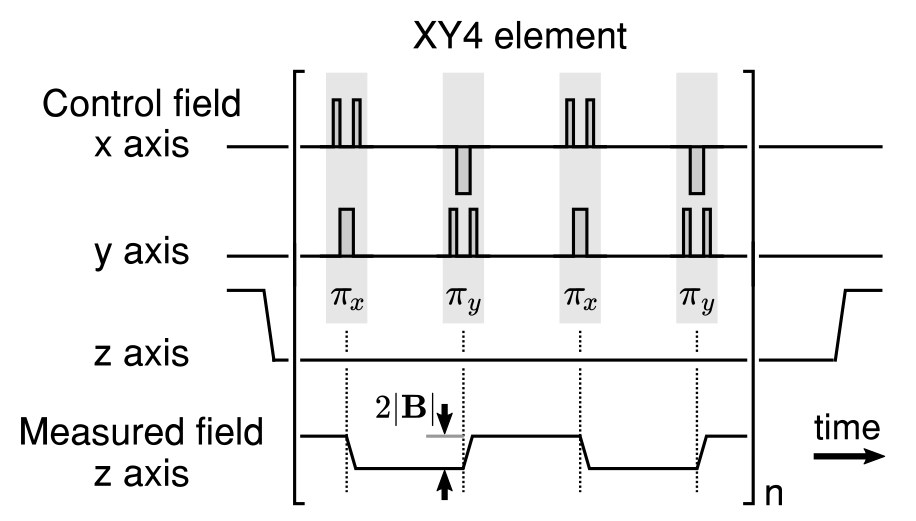}%
\caption{Experimental sequence.  Upon field switching to near-zero fields, alternating $(\pi)_x \equiv (\pi/2)_x(\pi)_y(\pi/2)_x$ and $(\pi)_y \equiv (\pi/2)_y(\pi)_{-x}(\pi/2)_y$ dc pulses (making the XY4 cycle) periodically invert the nuclear magnetization.  A magnetometer measures the $z$ field component throughout. 
}
\label{fig:figure1}\vspace{0.5cm}
\end{figure}

 
A suitable pulse sequence to detect $\mathbf{B}$, assumed parallel to the $z$ axis, is shown in \autoref{fig:figure1}.  Uniformly spaced $\pi$ spin flips toggle the sign of the $z$ component, producing a square wave that is distinct from background fields\cite{Ganssle2014ANIE}.  
Flips are applied as $\pi$ rotations alternately about $x$ and $y$ laboratory-frame axes making an XY4 cycle with repeating element ($\tau/2-\pi_{\rm X}-\tau-\pi_{\rm Y}-\tau-\pi_{\rm X}-\tau-\pi_{\rm Y}-\tau/2$).  The useful feature of XY4 is the dynamical decoupling of the background field from the spin system at the first-order average Hamiltonian level \cite{Bodenstedt2021jpclett}, which allows us to neglect bias fields with strengths $|\mathbf{B}_\mathrm{bias}| \ll 2\pi/\gamma\tau$.

\begin{figure}%
\includegraphics[width=0.75\columnwidth]{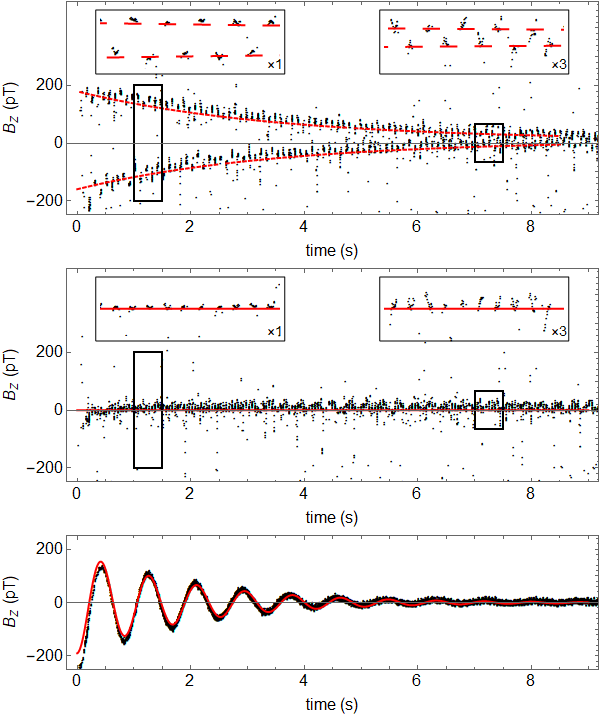}%
\caption{Measured $z$-axis component of magnetic field ($B_z$, relative to background) during the ultralow-field portion of the sequence shown in \autoref{fig:figure1}.  Single-shot signals are measured for \SI{1}{\centi\meter\cubed} of H\textsubscript{2}O using a \textsuperscript{87}Rb-vapor OPM under the following conditions: (top) XY4 decoupling, sample prepolarized at \SI{2}{\tesla}, (middle) XY4 decoupling, prepolarized at Earth's field, (bottom) free precession, i.e., without XY4 pulses, prepolarized at \SI{2}{\tesla}.  The red curves indicate fitting to square- or sine-modulated monoexponential decay curves with the time constants (top) $T_1 = \SI{3.5}{\second}$ and (bottom) $T_2^\ast = \SI{2.0}{\second}$. }%
\label{fig:figure2}
\end{figure}

The top panel of \autoref{fig:figure2} shows the NMR signal recorded during XY4 ($\tau=50$ ms, 0.2 ms $\pi$-pulse length) for a 1 \si{\centi\meter\cubed} cylindrical volume of 2-T-prepolarized water, using an alkali-metal-vapor OPM inside a magnetic shield (MS-1LF, Twinleaf LLC).  The OPM measures the $z$-axis component, $B_z$, of total field using quadrature demodulation of the Hanle absorption resonance in a \textsuperscript{87}Rb vapor\cite{Shah2007}.  The vapor is confined to a microelectromechanical system  (MEMS) cell of approximate dimensions 4 mm $\times$ 4 mm $\times$ 1.5 mm at a standoff distance of 3 mm from the sample chamber.  
The magnetic shield provides a simple and convenient way to isolate the detector from unknown and/or time-dependent background magnetic fields. The shield presents no additional operating constraints; indeed, it allows the apparatus to be carried freely around the room, and placed within a few meters of an MRI scanner in background fields of several tens of mT without any need to compensate for background drifts.
For further technical details of the OPM we refer the reader to the Supporting Information, plus past work\cite{Tayler2022PRApplied}.  The signal shown is processed minimally, where data points during each $\pi$ pulse and up to 5 ms afterwards are excluded (where the magnetometer output saturates) and mains-hum noise is suppressed by applying a 45-to-55-Hz band-stop filter.  The solid red curve: $B_z(t) = (190\,\,\rm pT)\times{\rm sgn}(\sin 20 \pi t)\exp(-t/T_1)$ fits to the remaining data points.  The decay time constant, $T_1=3.5$ s, is that of longitudinal relaxation under dynamical decoupling in the effective zero field, inclusive of the contribution due to imperfections in the $\pi$ pulses.

The middle panel of \autoref{fig:figure2} shows the OPM signal when the starting magnetic field is much lower, \SI{50}{\micro\tesla} or around Earth's field, where there is no major alternation seen above the noise background. As a comparison, the lower panel of \autoref{fig:figure2} shows the signal for 2-T-prepolarized water in the absence of XY4 decoupling.  Here the free-decay signal is an exponentially decaying cosine wave $B_z(t) = (190\,\,\rm pT)\times\cos (2.48 \pi t)\exp(-t/T_2^\ast)$ with $T_2^\ast = \SI{2.0}{\second}$, corresponding to precession in the approximately \SI{30}{\nano\tesla} remanent field of the magnetic shield.  The precession signal undergoes faster decoherence ($T_2^\ast<T_1$) due to gradients in the precession field, which are otherwise decoupled under XY4. 

\begin{figure*}%
\includegraphics[width=\textwidth]{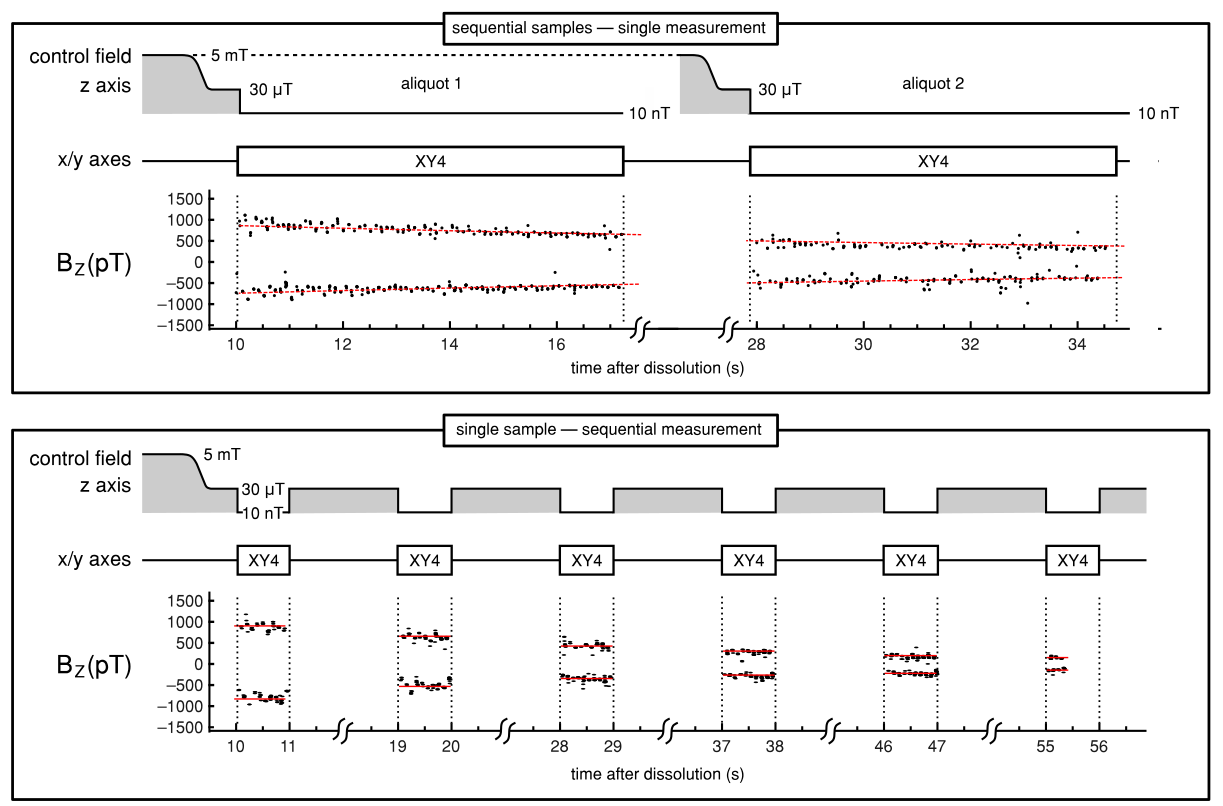}%
\caption{The NMR signals recorded for \SI{0.08}{\molar} hyperpolarized [1-\textsuperscript{13}{C}]-pyruvate under \textsuperscript{13}{C} XY4, $\tau=50$ ms. Upper panel $B_z$ data are obtained after portioning the hyperpolarized solution into two \SI{1.0}{\centi\meter\cubed} aliquots, each stored at 5 mT and then placed sequentially next to the zero-field OPM.  Red curves show monoexponential decays with fitted decay time constant $T_1(\mathrm{zero\, field}) = \SI{24+-1}{\second}$.  The lower panel data were obtained with a single aliquot, by alternating 1-s acquisition periods at zero field with 8-s storage periods at \SI{30}{\micro\tesla}. Signal loss in between acquisition periods is attributed to natural decay, fitted by the red curve to $T_1(\SI{30}{\micro\tesla}) = \SI{24+-1}{\second}$, which within error is equal to $T_1(\mathrm{zero\, field})$.}%
\label{fig:figure3}
\end{figure*}

To apply XY4 to a system of hyperpolarized \textsuperscript{13}C spins, the duration of each dc pulse is increased by a factor $\gamma_{\rm H}/\gamma_{\rm C}\approx 3.97$ to produce rotation angles of $\pi$ and $\sim$$4\pi$ on \textsuperscript{13}C and \textsuperscript{1}H, respectively.  In addition to dynamically decoupling small residual magnetic fields, this combination of angles also decouples the \textsuperscript{13}C-\textsuperscript{1}H scalar ($J_{\rm{CH}}$) coupling interaction when performed sufficiently quickly; $\tau\ll |1/J_{\rm CH}|$. The polarimetry technique can therefore be applied across a wide range of organic molecules. 

To demonstrate application to substances of clinical and pre-clinical relevance, we measured the OPM signal during \textsuperscript{13}C XY4 dynamical decoupling for solutions of hyperpolarized [1-\textsuperscript{13}C]-pyruvic acid (PA = \textsuperscript{12}CH\textsubscript{3}{\textsuperscript{12}C}O\textsuperscript{13}COOH; $^3J_{\rm CH}=1.3$ Hz). The samples were prepared initially as \SI{24}{\micro\liter} volumes of 0.2 wt.$\%$ OX-063 radical (GE Healthcare) and 0.09 wt.$\%$ Dotarem (Guerbet S.A., France) dissolved in neat-liquid PA (Sigma Aldrich).  Each of these were polarized at \SI{1.4}{\kelvin} in a magnetic field of \SI{3.5}{\tesla}, using a commercial dissolution-DNP polarizer (HyperSense, Oxford Instruments Ltd.) via 40 minutes of positive-lobe microwave irradiation near 95 GHz.  After polarization, the cryogenic samples were dissolved in a superheated phosphate-buffered-saline solution (4.5 \si{\centi\meter\cubed} in \textsuperscript{1}H\textsubscript{2}O, \SI{10}{\milli\molar} Na\textsubscript{2}HPO\textsubscript{4}, \SI{2}{\milli\molar} KH\textsubscript{2}PO\textsubscript{4}, 1 wt.$\%$ HEPES, 0.01 wt.$\%$ EDTA, 0.1 wt.$\%$ NaCl, 0.2 wt.$\%$ NaOH) and ejected from the instrument at a temperature of \SI{37}{\celsius} into a vial, at a final concentration of $C=0.080(1)$ \si{\molar}.  

Immediately after the dissolution process, two \SI{1.0}{\centi\meter\cubed} aliquots of the polarized \textsuperscript{13}C sample were collected and stored in the approximately \SI{5}{mT} fringe magnetic field of the DNP polarizer.  The procedure simulates a situation where fractions of a hyperpolarized sample are inserted into an MRI scanner one after the other\cite{Lees2020NMRBiomedicine}.  The aliquots were then transferred sequentially by syringe into the magnetic shield located approximately 1 m from the polarizer and containing the OPM and sample chamber used for the 2-T-prepolarized water sample.  A solenoid coil provided a guiding magnetic field\cite{Savukov2005PRL} of \SI{30}{\micro\tesla} over the syringe during transport.  For the first aliquot, the guiding field was switched off around 10 seconds after dissolution and the XY4 sequence ($\tau=50$ ms) was initiated and sustained for a time of 6-7 seconds. The liquid was then removed from the apparatus by syringe, the guiding field was switched on again, and the process repeated for the second aliquot; see \autoref{fig:figure3}, upper panel.  During the time at zero field, the OPM signal was recorded, also shown in the upper panel of \autoref{fig:figure3}.  Signals were fit by a monoexponential decay envelope with time constant $T_1 = 24(1)$ s.  The time constant is remarkably long, considering that the paramagnetic Dotarem (Gd\textsuperscript{3+}) and OX-063 species were not removed from the solution. This indicates a low seriousness of paramagnetism-induced relaxation in the zero-to-ultralow-field regime\cite{Bodenstedt2021natcomm}. 
One explanation could be that although Gd\textsuperscript{3+} ions remains in the liquid after dissolution, their contact with \textsuperscript{13}C spins is minimized due to chelation in the buffered solution.  The initial amplitude of the \textsuperscript{13}C signal, furthermore, is used to provide a lower-bound value for the polarization.  Amplitudes of the signals for the first and second pyruvate aliquots, 1020 pT and 600 pT respectively, at zero decoupling time equate via \autoref{eq:PA} to \textsuperscript{13}C polarizations of $P=$ 12(1) \% and 7(1) \%, using the 190 pT reference signal amplitude for the 2-T-prepolarized water sample (\autoref{fig:figure2}) and the other parameters as given above. The first one is in agreement with the \textit{ex post} value $P=$ 13\% calculated from the hyperpolarized vs.\ thermal \textsuperscript{13}C-NMR spectra of a third 1.0 \si{\centi\meter\cubed} aliquot from the same batch, which was also injected at 10.0 s into a bench-top 1.5 T NMR spectrometer (Pulsar, Oxford Instruments Ltd.). 

Sample polarimetry is not limited to one-off measurement.  As the lower part of \autoref{fig:figure3} shows, hyperpolarized spins may be sustained across multiple zero-field measurements performed sequentially.  After sample injection into the magnetic shield, the field is repeatedly and rapidly ($<$10 \si{\micro\second}) toggled between the effective-zero field and a \SI{30}{\micro\tesla} holding field to emulate polarization storage in ambient magnetic field.  In the case shown, 8-s storage intervals occur between 1-s zero-field observation windows.  For [\textsuperscript{13}C]-pyruvate, signal decay occurring between windows can be fit to the monoexponential time constant $T_1 = 24$ s, which within error is the same as at zero field under XY4 decoupling. 
The observation that the \textsuperscript{13}C $T_1$ is unchanged across three orders of magnitude in field is not so surprising for such a small molecule, given that paramagnetic relaxivities in aqueous solution are generally found to be constant below \SI{1}{\milli\tesla} \cite{Bodenstedt2021natcomm}.  We expect that the use of nonpersistent-radical polarizing agents\cite{Eichhorn2013PNAS110,Pinon2020CommChem3,Gaunt2021ANIE,Negroni2022JACS}, filtration\cite{Gajan2014PNAS} or other strategies to remove paramagnetic dopants from the solution should prolong the decay time at both the storage and detection fields.  

Given the growing interest in hyperpolarized MRI, a wide scope of application for the method is envisioned. Solid hyperpolarized [\textsuperscript{13}C]-pyruvate can be transported far away from where it is produced \cite{Ji2017,Capozzi2021CommChem}, for instance to an MRI facility at a hospital located hundreds of kilometers away from the polarizer\cite{Capozzi2022}.  
Nondestructive polarization measurement at the use end using a combined dissolution-and-magnetometry device may form part of the quality control procedure carried out upon liquification.  
Unlike conventional NMR, our technique does not rely upon maintaining a homogeneous magnetic field around the sample; this means that mandatory procedures for human in-vivo application including sample concentration---via absorption spectroscopy---and pH checks, could be measured in parallel.  This may significantly reduce the delivery time into a human body, which currently stands around 1 minute after dissolution and incurs a substantial natural decay of the polarization\cite{Kurhanewicz2019Neoplasia}.
Hyperpolarization procedures involving chemical reactions of parahydrogen at \si{\micro\tesla} fields\cite{Knecht2021PNAS,Marshall2022arxiv} should also be compatible with detection via OPMs, enabling real-time in-situ monitoring (or optimization) of polarization buildup. 

Most liquid-state hyperpolarized NMR and MRI procedures focus on the study of specific chemical sites in isotopically pure compounds (for instance, the terminal carboxyl carbon in pure [1-\textsuperscript{13}C]-PA).  However, it is possible, although uncommon, to hyperpolarize and study more than one isotopomer or compound at the same time, e.g., a solution of [1-\textsuperscript{13}C]-pyruvate and [1-\textsuperscript{13}C]-butyrate\cite{Bastiaansen2016SciRep} or [\textsuperscript{13}C]-urea\cite{Qin2021MRM87}.  The present version of our experiment does not allow polarimetry of two alike spin species at the same time and therefore the magnetometry procedure would have to be performed on the component pure solutions before mixing.  In principle, isotopomer-selective $\pi$ pulses that exploit the chemically specific Zeeman shifts and/or spin-spin J couplings could be used to achieve selectivity in mixtures.  Various other schemes may also be possible, such as magnetic gradiometry of two chemically identical samples that have opposite polarization in one of the compounds.

As a final remark, the square-wave NMR signal may in principle be recorded at any magnetic field, given a sufficiently sensitive detector.  In this work, the nT bias field was chosen to use a zero-field OPM of sensitivity around \SI{10}{\femto\tesla\per\sqrt\hertz} \cite{Tayler2022PRApplied}, although the technical noise background (such as mains hum noise) is much larger, on the order of \SI{1}{\pico\tesla\per\sqrt\hertz}.  Miniaturized versions of zero-field OPMs are produced on a medium commercial scale\cite{quspinQZFMGen3,twinleaf_microserf} for research use in magnetoencephalography\cite{tierney2019optically,Brookes2022TrendsNeurosci} and other biomagnetic imaging studies.  While these must operate in shielded environments for highest performance\cite{Savukov2020JMR,Blanchard2020JMR}, other types of OPMs and magnetic gradiometers can reach sensitivities of tens of \si{\femto\tesla\per\sqrt\hertz} at ambient field\cite{Limes2020} while also being less susceptible to background noise.  In the future, these may remove the current experimental requirement of a magnetic shield.    

\begin{acknowledgement}
The work described is funded by: 
EU H2020 Marie Sk{\l}odowska-Curie Actions project ITN ZULF-NMR  (Grant Agreement No. 766402); the European Union’s Horizon 2020 research and innovation program (GA-863037);
the Spanish Ministry of Science MCIN with funding from European Union NextGenerationEU (PRTR-C17.I1) and by Generalitat de Catalunya ``Severo Ochoa'' Center of Excellence CEX2019-000910-S; projects OCARINA (PGC2018-097056-B-I00), SAPONARIA (PID2021-123813NB-I00) and MARICHAS (PID2021-126059OA-I00),  funded by MCIN/ AEI /10.13039/501100011033/ FEDER ``A way to make Europe''; 
Generalitat de Catalunya through the CERCA program;  
Ag\`{e}ncia de Gesti\'{o} d'Ajuts Universitaris i de Recerca Grant Nos. 2017-SGR-1354 and 2021 FI\_B\_01039;  
Secretaria d'Universitats i Recerca del Departament d'Empresa i Coneixement de la Generalitat de Catalunya, co-funded by the European Union Regional Development Fund within the ERDF Operational Program of Catalunya (project QuantumCat, ref. 001-P-001644); 
Fundaci\'{o} Privada Cellex; 
Fundaci\'{o} Mir-Puig;
and the BIST--“la Caixa” initiative in Chemical Biology (CHEMBIO).
MCD Tayler and I Marco-Rius acknowledge financial support through the Junior Leader Postdoctoral Fellowship Programme from ``La Caixa'' Banking Foundation (projects LCF/BQ/PI19/11690021 and LCF/BQ/ PI18/11630020, respectively). 
\end{acknowledgement}

\begin{suppinfo}
\noindent Additional experimental details including photographs of the experimental setup (PDF).
\end{suppinfo}


\bibliography{references}

 \begin{figure*}
 \includegraphics[page = 1, width=\textwidth]{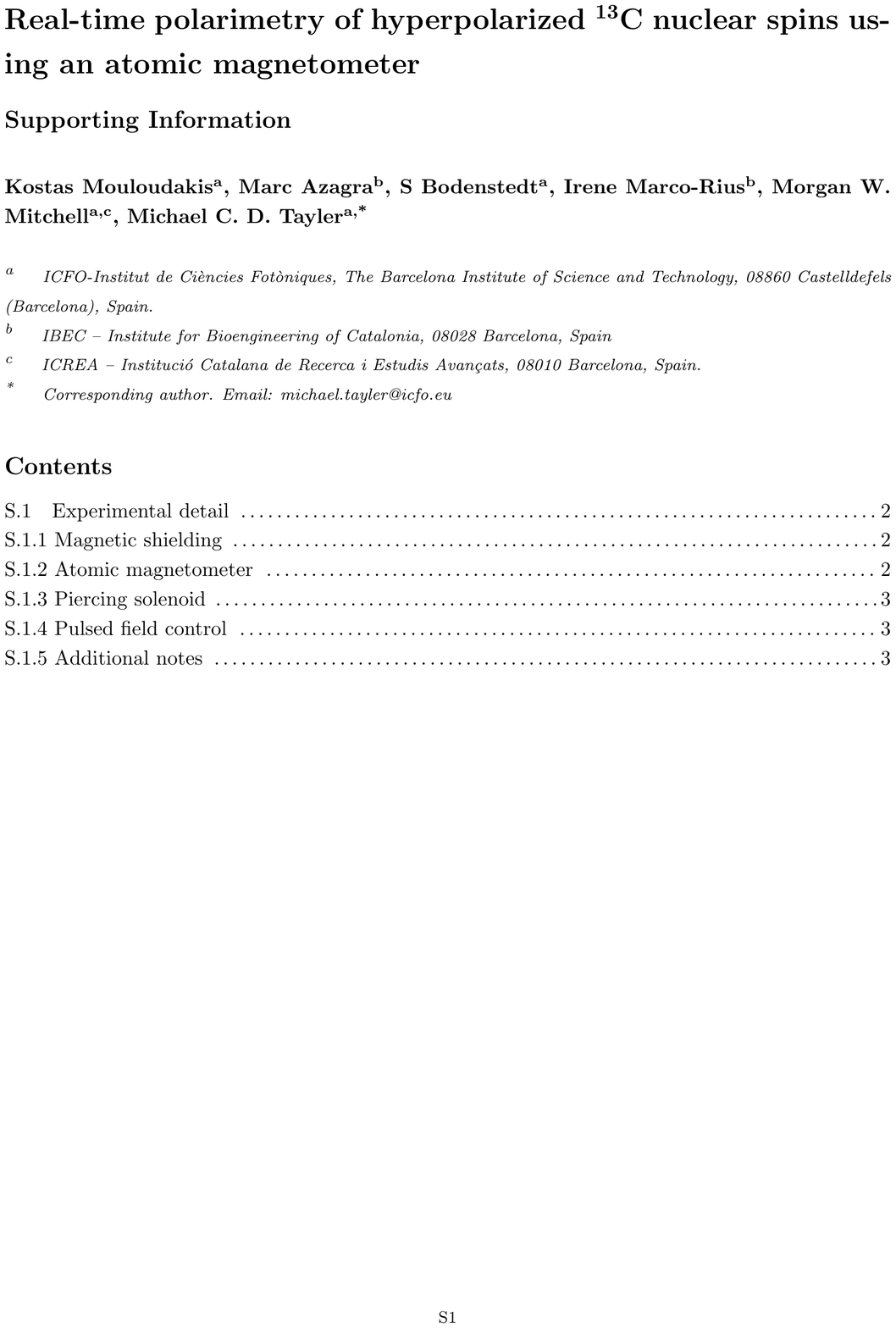}
 \end{figure*}

\begin{figure*}
\includegraphics[page = 2, width=\textwidth]{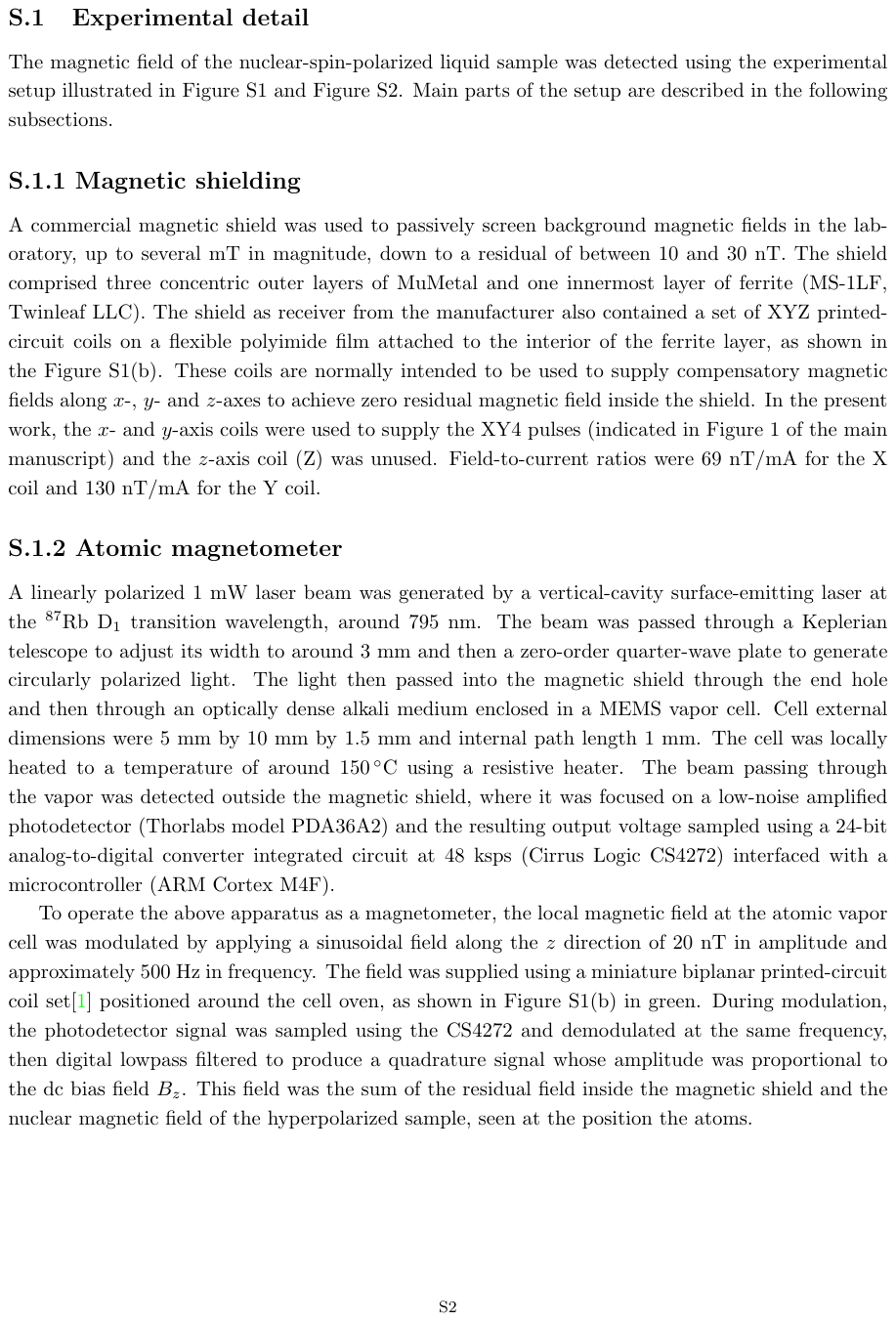}
\end{figure*}

\begin{figure*}
\includegraphics[page = 3, width=\textwidth]{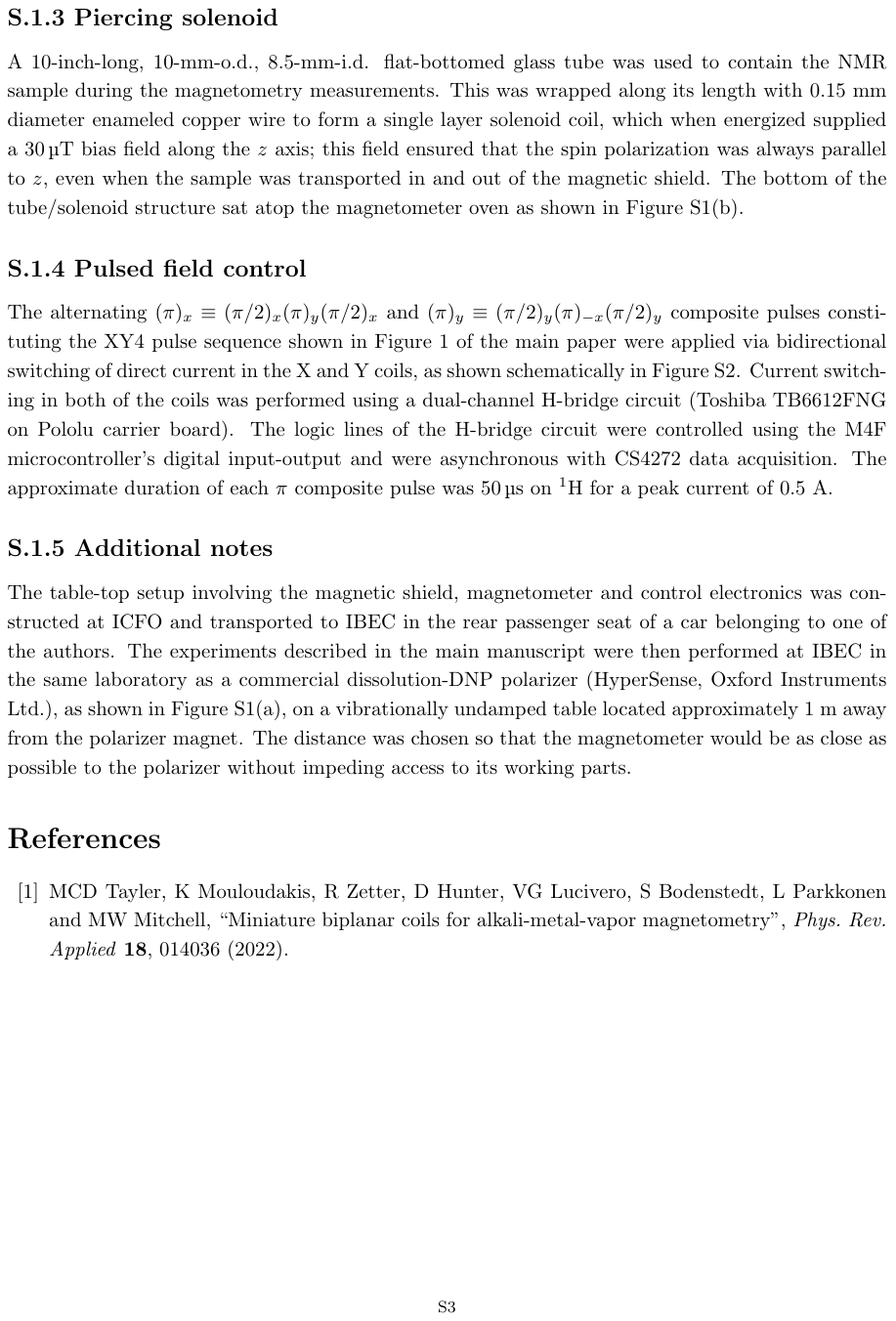}
\end{figure*}

\begin{figure*}
\includegraphics[page = 4, width=\textwidth]{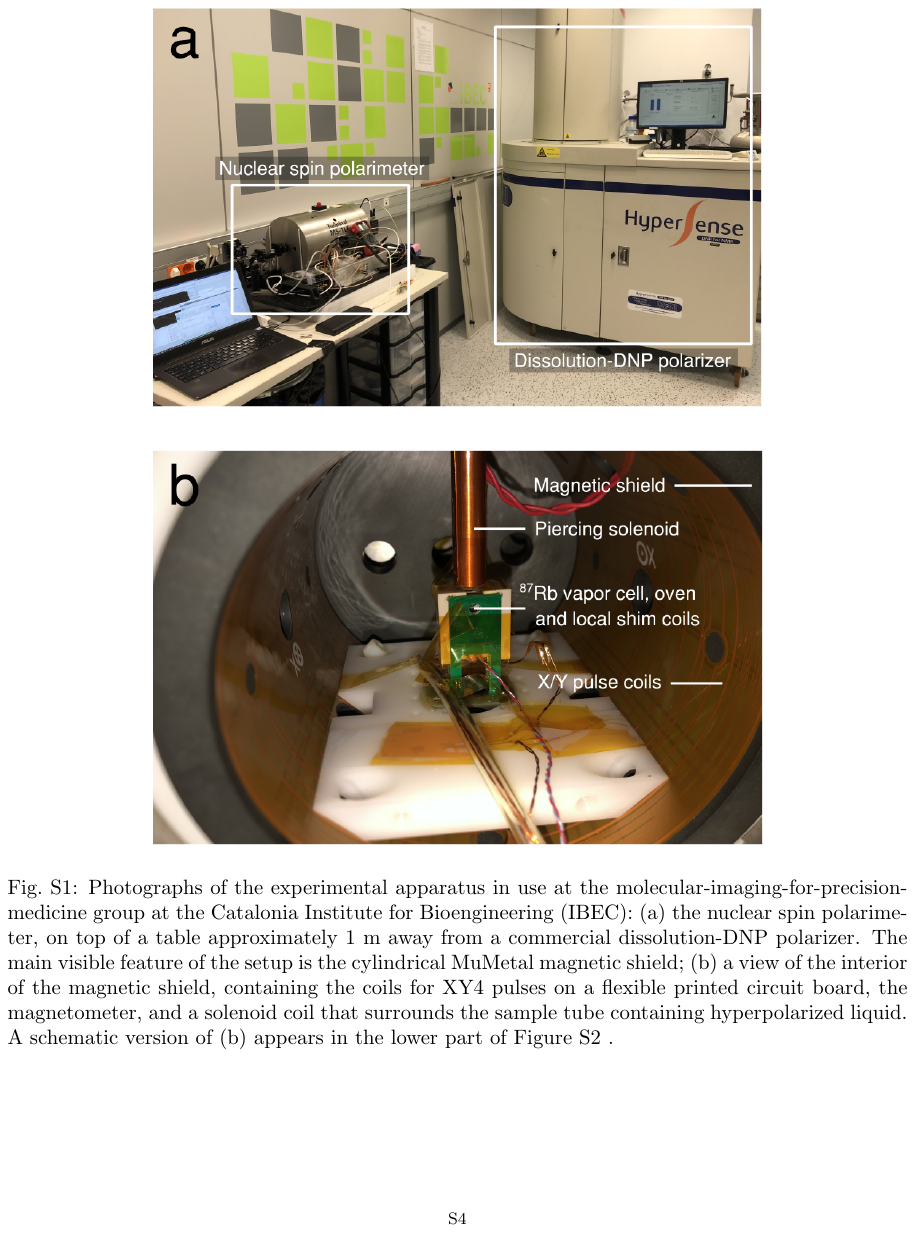}
\end{figure*}

\begin{figure*}
\includegraphics[page = 5, width=\textwidth]{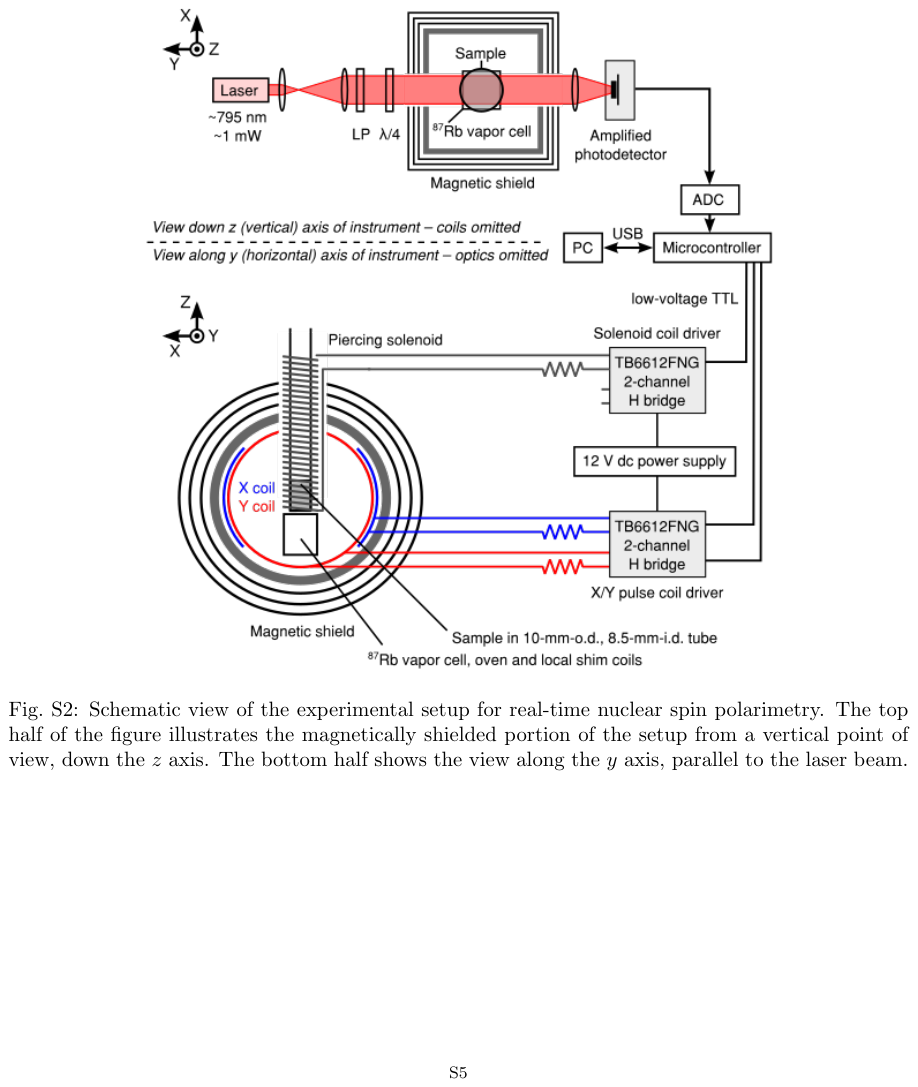}
\end{figure*}

\end{document}